\documentclass[final]{statsoc}
\pdfoutput=1

\usepackage[utf8]{inputenc}

\usepackage[a4paper]{geometry}
\usepackage{color}
 
\usepackage{graphicx}

\usepackage{xr}

\makeatletter
\newcommand*{\addFileDependency}[1]{
  \typeout{(#1)}
  \@addtofilelist{#1}
  \IfFileExists{#1}{}{\typeout{No file #1.}}
}
\makeatother

\newcommand*{\myexternaldocument}[1]{%
    \externaldocument{#1}%
    \addFileDependency{#1.tex}%
    \addFileDependency{#1.aux}%
}

\myexternaldocument{supp_mat}

\usepackage{hyperref}

\usepackage{etoolbox}

\makeatletter
\patchcmd{\@makecaption}
  {\parbox}
  {\advance\@tempdima-\fontdimen2} 
  {}{}
\makeatother  

\usepackage{amsmath, amssymb} 
\counterwithout{equation}{section}
\usepackage{bm}

\usepackage{natbib}

\title[CO\textsubscript{2} Projections Emulation]{Bayesian Functional Emulation of CO\textsubscript{2} Emissions on \\ Future Climate Change Scenarios}

\author{Luca Aiello} 
\address{Department of Mathematics, Politecnico di Milano, Italy.}
\address{Department of Economics, Management and Statistics, Università degli Studi di Milano Bicocca, Italy.}

\author{Matteo Fontana}
\address{MOX - Department of Mathematics, Politecnico di Milano, Italy}
\address{European Commission, Joint Research Centre (JRC), Ispra, Italy}

\author[Aiello, Fontana \& Guglielmi]{Alessandra Guglielmi}
\address{Department of Mathematics, Politecnico di Milano, Italy.}

\email{l.aiello4@campus.unimib.it}

\begin{document}

\begin{abstract}
We propose a statistical emulator for a climate-economy deterministic integrated assessment model ensemble, based on a functional regression framework. Inference on the unknown parameters is carried out through a mixed effects hierarchical model using a fully Bayesian framework with a prior distribution on the vector of all parameters. We also suggest an autoregressive parameterization of the covariance matrix of the error, with matching marginal prior. In this way, we allow for a functional framework for the discretized output of the simulators that allows their time continuous evaluation.
\end{abstract}


\section{Introduction}\label{sec:intro}

Climate change is, by far, the biggest and most life-threatening challenge of the present time. Its consequences have the potential to change our society and to threaten the very existence of humankind on this planet. Due to its complex nature, climate change is, at its core, a multidisciplinary problem. It involves many different disciplines, in the natural, human and social sciences, specifically physics, chemistry, engineering, economics and sociology. 

Due to this intrinsic multidisciplinary nature, and to properly simulate and understand the complexities in the interaction between the different systems that constitute our society, and climate, a special class of models has been devised. To analyze climate change, in the last years the scientific community has implemented several integrated assessment models (IAMs), which are deterministic computer models, i.e. simulators, that, given the same inputs, produce the same outputs every time a run is repeated. These models are complex simulations that are able to take into account the multidisciplinary nature of the problem, and consequently mix the numerous ingredients coming from different fields. These kind of large scale computer simulations are widely used in modern scientific research to investigate, among others,  physical phenomena that are too expensive or impossible to replicate directly \citep{fan2009numerical,textor2005numerical}. Often, the research interest is focused on quantifying how uncertainty in the input arguments propagates through the simulator and produce a distribution function over one or many outputs of interest, as well as how much uncertainty is introduced via the modelling effort.

Generating the same outputs with several IAMs (namely, creating a model ensemble) allows to quantify both parametric and model uncertainty. Parametric uncertainty refers to the variability induced by how much are we uncertain with respect to the right set of model input parameters, while model uncertainty refers to the ability of each IAM to model correctly only a part of the reality, and to the differences in the model implementation and modelling choices. It is of paramount importance to disentangle the key drivers of uncertainty in emissions projections because this understanding can help design better climate hedging strategies. Moreover, IAMs diagnostics is a relatively nascent field that is growing in importance to help validate these kind of computational models.

\begin{figure}[!htbp]
    \centering
    \includegraphics[width = 0.33\textwidth]{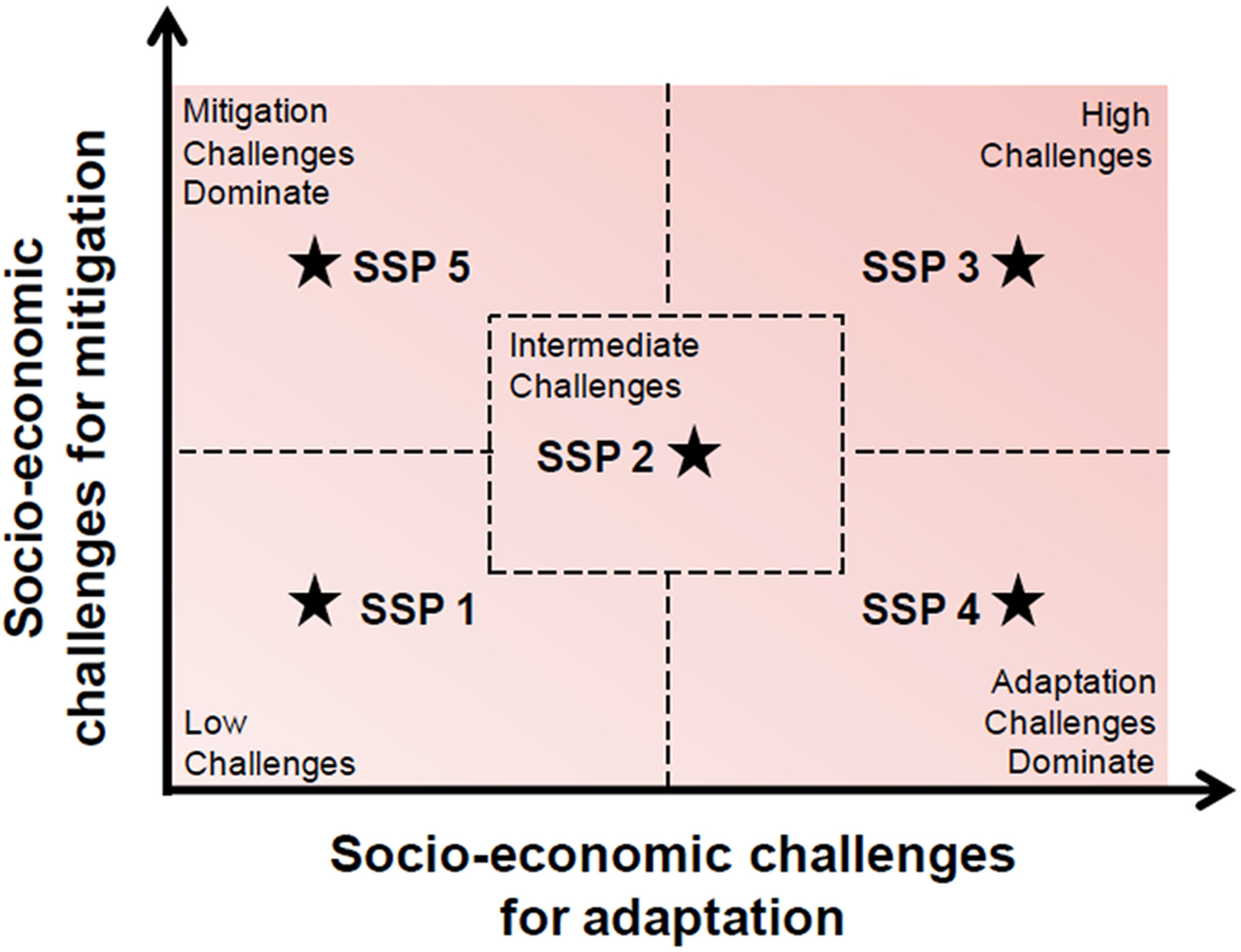}
    \caption{SSPs plane on adaptation and mitigation, based on Figure 1 from \cite{o2014new}}
    \label{fig:ssp_matrix}
\end{figure}

To properly address the fundamental parameter uncertainty in climate change modelling, the scientific community has adopted a scenario approach. Indeed IAMs use, as inputs, a set of so-called Shared Socio-economic Pathways (SSPs). Scenarios showing future greenhouse gas emissions are needed to estimate climate impacts and mitigation efforts required for climate stabilization. SSPs are part of a new framework that the climate change research community has adopted to facilitate the integrated analysis of future climate impacts, vulnerabilities, adaptation, and mitigation. Further information about the scenario process and the SSP framework can be found in \cite{moss2010next}, \cite{van2014new}, \cite{o2014new}, \cite{kriegler2014new}. An SSP consists in the discretization of a continuous plane of mitigation (i.e. reducing the modification in climate induced by human activitiy) and adaptation (i.e. how we adapt to the changing climate) to climate change \citep[][]{riahi2017shared} as in Figure~\ref{fig:ssp_matrix}.

5 SSP scenarios have been identified. Three of them belong to the main diagonal, describing futures where the challenges for both climate change adaptation and mitigation are low (SSP1), intermediate (SSP2) and high (SSP3).  In addition there are two asymmetric scenarios that do not belong to the main diagonal. In fact SSP4 has high challenges for adaptation combined with low challenges for mitigation, while the contrary holds for SSP5. We focus on the same application as in \cite{marangoni2017sensitivity}, from which we take the data. As in \cite{marangoni2017sensitivity} we use only the SSP belonging to the main diagonal of the matrix represented in Figure~\ref{fig:ssp_matrix}.

The IAMs output consist in projections of several variables, and the one that we consider here are the carbon dioxide emissions, since this is the most studied variable affecting the climate. For this reason data consist of $23 \times 5$ CO\textsubscript{2} global emission (expressed in GtCO\textsubscript{2}) profiles. Each CO\textsubscript{2} profile is time dependent, discretized with a ten-year frequency, to ease the computational burden of the IAMs. In fact, these massive simulations are expensive and demanding, and a higher frequency in the output could increase the general cost of the computation, as well as the time needed to run the model. Those emissions have been computed as the output of 5 different IAMs, using as input the same combinations of SSP variables, such as gross domestic product and fossil fuel availability.

Learning from \cite{o2006bayesian}, an important feature of this context that is worth stressing is that, the output of a simulator (in our case an IAM), which is a computer prediction of the real phenomenon, will inevitably be imperfect. Statistical analysis should be able to incorporate the model biased representation of the real process. It is of great interest to have one single fast statistical emulator (i.e. a model emulating the unknown outputs from the simulator) able to capture all the variability induced by the simulators \citep{kennedy2001bayesian,santner2003design}. In particular, there are certain aspects of CO\textsubscript{2} simulation scenarios, that can never be known with certainty, since we cannot run each model for an infinite length of time or with an infinite number of possible starting values of the simulations. Part of the uncertainty is also due to the inability to run the model for every possible choice of the input parameters. The aforementioned issues are those inducing uncertainty on the data coming from the runs performed. The quantification of uncertainty in this context, performing a run with all the combinations of the inputs variable in order to learn the input-output map, could be too expensive. Conversely, choosing the parameters from a sparse grid could be useless in learning key features. Ultimately, the weaknesses of this approach mainly originates by the model simplifications.

Here by statistical emulator we mean a statistical model representing data which are the output of different IAMs. For the aforementioned issues concerning huge computer simulations, this work proposes a statistical emulator, treating the IAMs models as black boxes in order to model the uncertainty in a non-intrusive way. An effective emulator is one that provides good approximations to the computer code output for wide ranges of input values, and accurate quantification of the emulation uncertainty \citep[see][]{francom2018sensitivity}. In fact, one of the biggest advantages of this approach is the possibility, after having emulated the process, of evaluating outputs with given inputs different from the one used to fit the model \citep{busby2009hierarchical}. Furthermore, the emulator will allow to have error intervals at unseen input, completing, in this way, basic requirements desired for uncertainty quantification. In simple words, an emulator is a stochastic representation of a computer model that generates a prediction for the output of a computer model at any setting of the model parameters and reports a measure of uncertainty for that prediction \citep{williamson2014evolving}. Here we follow the Bayesian approach, that is to assume all the parameters, in the statistical model, as random distributed according to a \textit{prior density}. One of the most recent works involving Bayesian emulation is \cite{francom2019inferring}, where they proposed a method based on adaptive splines in order to model simulations about the spread of radioactive particles through the atmosphere.

Mathematically speaking, a computer model is a function of a possibly large number of parameters. From a mathematical perspective IAMs are simulators that can be considered as functions. More precisely, given an input $\mathbf{x} \in \mathcal{X} \subseteq \mathbb{R}^p$, a simulator is a function $f:\mathcal{X} \mapsto \mathbb{R}^q$ such that $\mathbf{y} = f(\mathbf{x})$, with the output represented by $\mathbf{y} \in \mathbb{R}^q$ \citep{conti2010bayesian}. Because of the high complexity of the computer model, $\mathbf{f}$ is taken as a black box; hence proper statistical modelling assumptions will be needed in order to estimate it.

To statistically emulate these deterministic simulators, the framework in which we decide to set the problem is \textit{Function-on-Scalar Regression} (FOSR), a context that is the extension to functional responses of the classic linear regression, where the response can be either a scalar or a vector. The functional approach is a useful way to represent easily the temporal dependence, and making inference or predicting in between the time units (in our case decades). For more details see \cite{ramsay2005functional}. Applications of FOSR are wide and interesting: examples include blood pressure profiles during pregnancy \citep{montagna2012bayesian}, longitudinal genome-wide association studies \citep{barber2017function, fan2017high} and analysis of actigraphy data to investigate the association between intraday physical activity and responses to a sleep questionnaire \citep{kowal2020bayesian}. FOSR has many features in common with multivariate regression - estimation and inference for the regression coefficients, and prediction of new responses - carrying on additional modeling challenges. Within-function dependence of functional data requires careful modeling of the covariance structure, which may be complex or require further assumptions (i.e. a parametric structure), with implications for model flexibility, computational complexity and scalability. Moreover, the regression coefficients in FOSR are function themselves, which complicates estimation, inference and interpretation.

As we said and motivated before, here we assume a Bayesian approach for our analysis and data, which are CO\textsubscript{2} emission profiles. We use a Bayesian multilevel hierarchical model, estimating the coefficients of the basis expansions of the regression coefficients, a method from FDA that allow to represent a function through some basis. Bayesian random effects models, also known as hierarchical models, for longitudinal data are very popular. Relevant references on the topic are \cite{daniels2002bayesian} and \cite{shen2020bayesian}. The model we assume is similar to the one proposed by \cite{goldsmith2016assessing} where the authors focused on functional representation of the trajectories of the arm of patients affected by stroke. The main difference we introduce is the form of the parametric covariance structure for the within-function correlation for computational ease and predictive purposes. The Bayesian approach to statistical inference has several benefits, such as the ability to incorporate multiple sources of information and uncertainty of the parameters, and a greater flexibility to build complex model structures. Concerning hierarchical modeling, it is a natural framework in which different kind of grouped data can be described, such as multilevel data of many subjects, as in our application. In our case the group will be the combination of SSPs input for the IAMs, because we are interested in assessing how the input of the simulations affects the outputs. In the Bayesian hierarchical models the usual prior distribution for all the group specific parameters is such that it allows borrowing information from the largest groups in order to give better estimates of parameters from the smallest ones. This type of Bayesian model is particularly appropriate in our application, since it allows to discover how scenarios are affecting the gas emissions during this century and more deeply what is the contribution of the various IAMs and SSPs. Moreover our approach is fully Bayesian, meaning that we assume a joint prior for all models unknown parameters and we compute the relative posterior inference, describing the new uncertainty of the parameters \lq\lq after having seen the data\rq\rq. One of the advantage of the Bayesian approach is the probabilistic interpretation of the results obtained. In fact, through the Bayesian approach we do not obtain punctual estimates but rather an entire distribution estimate, easing in this way the assessment of uncertainty quantification through the credibility intervals. One of the main advantages of this approach, consists in providing joint inference on parameters and predicted variables. This is quite a novelty in the Climatic simulations framework, since past works in literature mainly focused on frequentist Functional Data Analysis.

We propose a five-fold original contribution in the present work
\begin{itemize}
    \item The proposal of a Bayesian model to make inference for CO\textsubscript{2} simulated emissions, an approach that is new for this type of simulator data;
    \item the choice of the Bayesian model (i.e. likelihood and prior)
    \item the understanding of the results we have obtained through a probabilistic perspective. In particular we have given a probabilistic measure of how much in the IAMs each SSP contributes to the output. Moreover we have written Stan \citep{rstan} and R \citep{R} codes to run the MCMC simulations for computing the posterior inference.
    \item modeling the variability of the single SSPs combinations induced by the different IAMs, giving a credible interval for a new IAM simulator;
    \item giving a continuous framework to the discretized output of the simulations, as to be able to evaluate emission at any time other than the one ouput by the simulation.
\end{itemize}

This manuscript is organized as follows: Section~\ref{sec:data} describes the data we consider in our analysis, Section~\ref{sec:BFOSR} the Bayesian model we have considered. In Section~\ref{sec:application} we show the application of our approach to the data and the findings. The article concludes with a discussion and further developments in Section~\ref{sec:conclusions}. In the end, Supplementary Material contains some details on hyper-parameters choice, sensitivity analysis and complementary images that were omitted from the main manuscript. 

\section{Data}\label{sec:data}

Data were taken from the work by \cite{marangoni2017sensitivity}. Further information and details on SSPs are available at the \href{https://tntcat.iiasa.ac.at/SspDb/dsd?Action=htmlpage&page=10}{International Institute for Applied System Analysis (IIASA) site}. We consider the output of 5 IAMs. Each IAM outputs 23 different time series, corresponding to 23 different inputs, with a ten year frequency, of the CO\textsubscript{2} global emission (expressed in gigatonnes of carbon dioxide - GtCO\textsubscript{2}). More precisely, an input given to an IAM is a combination of future projection scenarios concerning the following SSP variables: population (\textbf{POP}), gross domestic product per capita (\textbf{GDPPC}), energy intensity improvements (\textbf{END}), fossil fuel availability (\textbf{FF}), low-carbon energy technology development (\textbf{LC}). In each of the 23 combination the scenarios describing the variables belong to one of the three levels described in Section \ref{sec:intro}: SSP1, SSP2, SSP3. While SSP variables do represent temporal pathways of given numerical inputs for an IAM, we adapt the same experimental strategy as in \cite{marangoni2017sensitivity}, thus mapping a given pathway to a specific level of the corresponding SSP variable. In other words, for our purposes, SSP variables are not functions, but scalars. Recall that the SSP plane is continuous (see Figure~\ref{fig:ssp_matrix}), and the SSP levels are taken from the discretized version of that plane (as explained in Section \ref{sec:intro}). Their representation consists in scalar variables indicating to which scenario the variable belongs. For example the combination in which \textbf{GDPPC} takes value 1 and all the other variables takes value 2 means that \textbf{GDPPC} in that case follows the SSP1 pathway and all the others the SSP2 one. The choice of treating the variables as continuous instead of discrete-valued, is principally motivated by the chance of making prediction on scenarios that are, for example, half-way between two levels (i.e. $\mathbf{GDPPC} = 2.5$).

The data we consider have the following structure: 
\begin{align*}
        \mathbf{y}_{ij} = \bigl[y_{ij}(t_1),y_{ij}(t_2),...,y_{ij}(t_D)\bigr] && \mathbf{w}_i = \bigl[1,w_{i1},...,w_{ip}\bigr]^T && j = 1,...,J, && i = 1,...,I
\end{align*}
where $J=5$ and $I=23$ are, respectively, the number of IAMs and SSPs combinations, $\{t_1,t_2,...,t_D\} = \{2020,2030,...,2090\}$ with $D=8$. Here $\mathbf{w}_i$ represent the $i$-th combination of SSP variables, including an intercept term, with $p=5$; $\mathbf{y}_{ij}$ is the CO\textsubscript{2} profile emission, in logarithmic scale, produced by the $i$-th SSP combination within the $j$-th IAM. The total number of data is $N = I \times J$.  Each of the five panels in Figure~\ref{fig:models_t_ser} in the Supplementary Material, associated to a different IAM, shows 23 different emission profiles. The plots help understanding how different IAMs produce very different projection ranges across time. This is principally due to the differences in the assumptions and in the implementation of the IAMs, and we are going to model those differences in our modeling approach.

\begin{figure}[!htbp]
    \centering
    \includegraphics[width=0.95\textwidth]{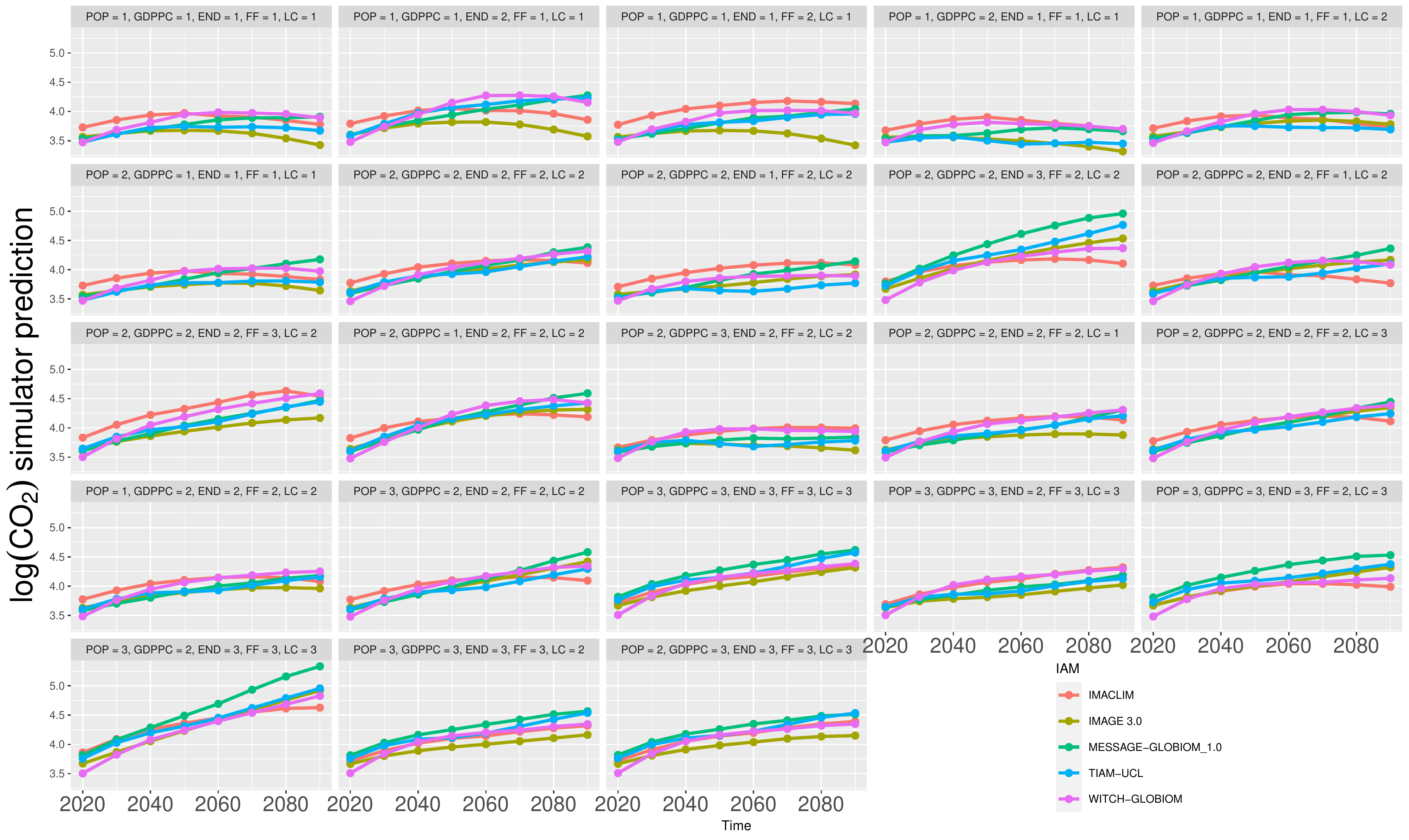}
    \caption{CO\textsubscript{2} emissions profiles in logarithmic scale grouped by SSP combination; each panel displays the 5 output from the same SSP combination.}
    \label{fig:SSP_t_ser}
\end{figure}

Figure~\ref{fig:SSP_t_ser} shows, for each of the 23 combinations of the SSP variables, the 5 curve corresponding to different IAMs. Within each panel, it is clear that the associated IAM curves are correlated. For this reason in the next section we will model this dependence through random effects.

\begin{figure}[!htbp]
    \centering
    \includegraphics[width=0.75\textwidth]{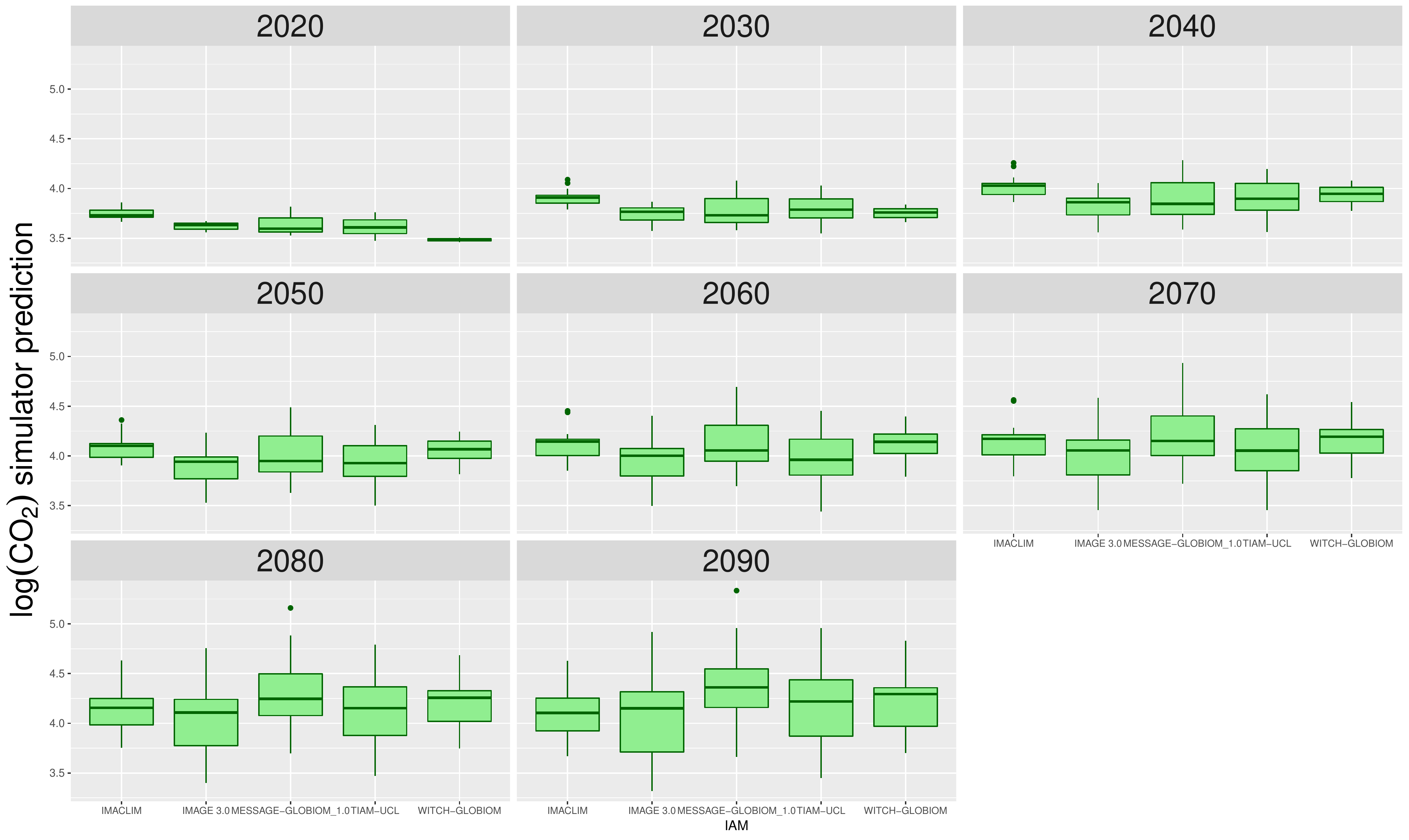}
    \caption{Boxplots of CO\textsubscript{2} emissions (in the log scale) by IAMs for each decade. The range of the y-axis is the same.}
    \label{fig:mod_box_each}
\end{figure}

Figure~\ref{fig:mod_box_each} shows box-plots of the emissions per decades comparing the output of the five IAMs. 
Note that for each decade the boxplots are different across the IAMs.

\section{Bayesian Functional Model}\label{sec:BFOSR}

In Section \ref{sec:data} we have introduced notation for our data, i.e. $y_{ij}(t_d)$ as the logarithm of the CO\textsubscript{2} emission at time $t_d$ produced by the $j$-th IAM given the $i$-th SSP combination as input, and $\mathbf{w}_{i} = \bigl[1,w_{i1},...,w_{ip}\bigr]^T$ as the vector describing the $i$-th SSP combination, with $p=5$. We assume, for any fixed $t$, for each SSPs combination $i=1,...,I$ and for each IAM $j=1,...,J$:
\begin{equation}\label{eq:time_model}
    y_{ij}(t) = \mathbf{w}_{i}^T \bm{\beta}(t) + c_{i}(t) + \epsilon_{ij}(t)
\end{equation}
where $\epsilon_{ij}(t)$ is a Gaussian random error with $0$ mean and $\bm{\beta}(t) = \bigl[\beta_{0}(t),\beta_{1}(t),...,\beta_{p}(t)\bigr]$ is the corresponding regression parameter at time $t$. Here $c_{i}(t)$ is the $i$-th SSP combination specific random effect coefficient function, that models the internal variability of the combination and induce correlation between observations with the same SSPs combination as input.

We rely on functional representation of the data since we assume the CO\textsubscript{2} emission simulations as a continuous phenomenon in time, that we can reasonably assume to be smooth, and whose temporal downscaling is of great interest. Here, by temporal downscaling, we mean the ability to assess and evaluate the output of the IAM for any time instant, and not only for those at which the IAMs output were produced. Recall that the IAMs, for a matter of computational cost, provide output only at the decades. For this reason, we believe that the functional approach could be very useful in the temporal downscaling (see Section \ref{subsec:em_curve_est}). 

Hence we consider a FOSR model for the response curves $\{y_{ij}(t)\}$, expanding all the regression parameters $\bigl\{\beta_{k}(t),k=0,1,...,p\bigr\}$ and random effects $\bigl\{c_{i}(t),i=1,...,I\bigr\}$ via a truncated B-spline basis expansion with $K$ components \citep[for more details, see][]{ramsay2005functional}. More explicitly, we assume that
\begin{align}\label{eq:beta_exp}
    \beta_{k}(t) = \sum_{l=1}^{K} b_{lk} \theta_{l}(t) && k = 0,1,...,p
\end{align}
where $\bigl\{\theta_{l}(t), l = 1,...,K\bigr\}$ is the B-spline basis and $\bigl\{b_{lk},l=1,...,K\bigr\}$ are the unknown scores of the functional parameters $\beta_{k}(t)$ expansion. Similarly we assume
\begin{align}\label{eq:c_exp}
    c_{i}(t) = \sum_{l=1}^{K} d_{li} \theta_{l}(t) && i = 1,...,I
\end{align}
where $\bigl\{d_{lk},l=1,...,K\bigr\}$ are the unknown scores of the functional random effect $c_{i}(t)$ expansion.

Eq. \eqref{eq:time_model} becomes, in a vectorized form, the following
\begin{equation}\label{eq:mix_eq}
    \mathbf{y}_{ij} = \mathbf{w}^T_i \bm{\beta} + \mathbf{z}^T_i \mathbf{c} + \bm{\epsilon}_{ij}, \text{\,\,\,\,\,\,\,\,\,\,\,\,} \bm{\epsilon}_{ij}|\bm{\Sigma} \stackrel{iid}{\sim} \mathcal{N}_D(\mathbf{0},\bm{\Sigma}) 
\end{equation}
where $\bm{\beta} = \bigl[\beta_{k}(t_d)\bigr]_{k,d}$ is the \textbf{unknown} $(p+1) \times D$ matrix whose rows are the regression coefficients functions evaluated at the time grid, $\mathbf{z}_i$ is the $I \times 1$  vector indicating the SSP scenario combination with all 0 elements but a 1 in the $i$-th position, $\mathbf{c} = \bigl[c_{i}(t_d)\bigr]_{i,d}$ is the \textbf{unknown} $I \times D$ matrix of subject random effects and $\bm{\epsilon}_{ij} = \bigl[\epsilon_{ij}(t_1),\epsilon_{ij}(t_2),...,\epsilon_{ij}(t_D)\bigr]^T$ is the $D$-dimensional error vector.

Expanding $\bigl\{\beta_{k}(t)\}$ and $\{c_{i}(t)\bigr\}$ as in \eqref{eq:beta_exp} and \eqref{eq:c_exp} respectively, the matrices $\bm{\beta}$ and $\mathbf{c}$ in \eqref{eq:mix_eq} assume the following form:
\begin{equation}\label{eq:mix_exp}
    \begin{aligned}
        \bm{\beta} = [\bm{\Theta}\mathbf{B}_W]^T && && && \mathbf{c} = [\bm{\Theta}\mathbf{B}_Z]^T
    \end{aligned}
\end{equation}
with $\bm{\Theta} = \bigl[\theta_{l}(t_d)\bigr]_{d,l}$ being the \textbf{known} $D \times K$ cubic B-splines evaluation matrix. Here $\mathbf{B}_W = \bigl[b_{l,k}\bigr]_{l,k}$ is the $K \times (p+1)$ matrix whose columns contain the fixed effects basis scores vector to be estimated and $\mathbf{B}_Z = \bigl[d_{l,i}\bigr]_{l,i}$ is the $K \times I$ matrix whose columns are the random effects basis scores vector to be estimated as well.

With the basis expansion described in \eqref{eq:mix_exp}, model \eqref{eq:mix_eq} is equivalent to the following: 
\begin{align}
    \mathbf{y}_{ij} = \mathbf{w}^T_i \mathbf{B}^{T}_W \bm{\Theta}^{T} + \mathbf{z}^T_i \mathbf{B}^{T}_Z  \bm{\Theta}^{T} + \bm{\epsilon}_{ij} && \bm{\epsilon}_{ij} \stackrel{iid}{\sim} \mathcal{N}_D(\mathbf{0},\bm{\Sigma}) 
\end{align}
for $i=1,...,I$ and $j=1,...,J$.

For a more efficient MCMC sampling performance, the model is hierarchically centered as in \cite{gelfand1995efficient}, that is, for $i=1,...,I$ and $j=1,...,J$:
\begin{equation}\label{eq:rand_eq}
    \mathbf{y}_{ij} = \mathbf{z}^T_i \mathbf{B}^{T}_Z  \bm{\Theta}^{T} + \bm{\epsilon}_{ij} \text{\,\,\,\, with \,\,\,\,} \bm{\epsilon}_{ij} \stackrel{iid}{\sim} \mathcal{N}_D(\mathbf{0},\bm{\Sigma}) 
\end{equation}
with centered marginal
\begin{equation}
    \begin{aligned}\label{eq:rand_prior}
    \mathbf{B}_{Z,i}|\mathbf{B}_{W},\sigma_Z^2 &\stackrel{ind}{\sim} \mathcal{N}_{K}(\mathbf{B}_W \mathbf{w}_i, \sigma_{Z}^2 \mathbf{P}^{-1}) & \sigma_{Z}^2 &\sim IG(a_{Z},b_{Z}) & i&=1,...,I \\[1em]
    \mathbf{B}_{W,k}|\sigma_{W,k}^2 &\stackrel{ind}{\sim} \mathcal{N}_{K}(\mathbf{0},\sigma_{W,k}^2 \mathbf{P}^{-1}) & \sigma_{W,k}^2 &\stackrel{ind}{\sim} IG(a_{W,k},b_{W,k}) & k&=0,1,...,p
    \end{aligned}
\end{equation}
where $a_{Z}$, $b_{Z}$, $a_{W,k}$ and $b_{W,k}$ are fixed hyperparameters. 

Model \eqref{eq:rand_eq}-\eqref{eq:rand_prior} is very similar to the model in \cite{goldsmith2016assessing}, but as it will be clear in a while, we introduce a different covariance structure on the error. In \eqref{eq:rand_prior} we assume the $K \times K$ matrix $\mathbf{P} = \alpha \mathbf{P}_0 + (1-\alpha) \mathbf{P}_2$ as a penalization matrix which is constructed as a weighted sum of two different matrices concerning shrinkage (i.e. $\mathbf{P}_0$) and smoothness (i.e. $\mathbf{P}_2$). Further details on the construction of the components of $\mathbf{P}$ are present in \cite{eilers1996flexible}. 

We assume an autoregressive configuration for the covariance matrix $\bm{\Sigma}$ of the errors

\begin{equation}\label{eq:ARH_cov}
    \bm{\Sigma} = 
    \begin{bmatrix} 
    \sigma^2 & \sigma^2\rho & \dots & \dots & \sigma^2\rho^{D-1} \\
    \sigma^2\rho & \ddots & & &
    \vdots &  \\
    \vdots & & \ddots & & \vdots \\
    \vdots & & & \ddots & \sigma^2\rho \\
    \sigma^2\rho^{D-1} & \dots & \dots & \sigma^2\rho & \sigma^2 
    \end{bmatrix}
\end{equation}
In a first version of this work \citep{aiello2020} we have assumed $\bm{\Sigma}$ as a full matrix and given it an IW prior distribution. However, posterior computations were much heavier and the associated inference gave evidence to the autoregressive assumption \eqref{eq:ARH_cov}.

We complete the prior specification assuming prior independence among $\sigma^2$, $\psi$ and $\rho$, and:
\begin{equation}\label{eq:arh1_prior}
    \begin{aligned}
        \sigma^2 \sim IG\biggl(\frac{\nu}{2},\frac{\nu}{2}\psi\biggr) && && && \psi \sim \Gamma\biggl(\frac{\nu_{0}}{2}, \frac{\nu_{0}}{2} \frac{1}{\psi_{0}}\biggr) && && && \rho \sim \mathcal{N}(0,1) 
    \end{aligned}
\end{equation}
where $\nu$, $\nu_{0}$ and $\psi_{0}$ are fixed hyperparameters.

Observe that \eqref{eq:ARH_cov} is a first order autoregressive covariance structure over time. The correlation between two successive decades is assumed homogeneous over time in \eqref{eq:ARH_cov} and denoted by $\rho$. This parameterization for $\bm{\Sigma}$ will result particularly convenient for computing predictions at times not included in the IAMs output.

Through model \eqref{eq:rand_eq}-\eqref{eq:arh1_prior} we reduced the infinite dimensional problem to a finite dimensional one.

\section{Application} \label{sec:application}

In this section we apply the model described in Section \ref{sec:BFOSR}, specifically \eqref{eq:rand_eq}-\eqref{eq:arh1_prior}, to data from the IAM model ensemble described in Section \ref{sec:data}. Remember that each SSP combination $\mathbf{w}_{i}$ is a $6$-dim vector input, including the intercept term, having multiple curves associated with it (one for each IAM).

We fix the number of B-spline basis to $K = 8$, and the adjusting parameter between shrinkage and smoothness to $\alpha = 0.01$ in $\mathbf{P} = \alpha \mathbf{P}_{0} + (1-\alpha) \mathbf{P}_{2}$ in order to give more weight to smoothness. We fixed $a_{W,k} = 4$ for all $k = 0,...,p$, $\mathbf{b}_{W} = [0.51,0.0002,0.002,0.001,0.0005,0.0001]^T$, $a_{Z} = 92$, $b_{Z} = 0.0038$ (see \eqref{eq:rand_prior} for definition of these hyperparameters), $\nu = 7$, $\nu_{0} = 2$ and $\psi_{0}=0.047$ (see \eqref{eq:arh1_prior}). See Section \ref{sec:hp_choice} in Supplementary Material for details on this choice.

Posterior inference is obtained through Stan \citep{rstan,hoffman2014no}, an open-source, general purpose programming language for Bayesian analysis, and rstan for interfacing with R \citep{R}. The codes are available in the Supplementary Material, Section \ref{sec:codes}.

Four chains were ran in parallel by Stan, each one with $20,000$ iterations, discarding the first $15,000$. Hence, the total final sample size is equal to $20,000$. Details on sensitivity analysis associated with hyperparameters and convergence diagnostics plots, can be found in the Supplementary Material, Section \ref{sec:hp_choice}.

\subsection{Regression coefficient functions}

We report here the estimate of the functional regression parameters $\bigl\{\beta_{k}(t),k=0,1,...,p\bigr\}$. More precisely, having access to the simulated posterior values of $\mathbf{B}_{W} = \bigl[b_{l,k}\bigr]_{l,k}$, the estimates of the regression coefficient functions can be computed as:
\begin{align}\label{eq:rand_est_coeffs}
    \mathbb{E}\bigl[\beta_{k}(t)|data\bigr] = \frac{1}{R} \sum_{r=1}^{R} \bm{\Theta}(t) \mathbf{B}_{W,k}^{(r)} && k=0,1,...,p
\end{align}
where R is the number of MCMC simulations (here $R=20000$), $\bm{\Theta}(t) = \bigl[\theta_{l}(t)\bigr]_{l}$ is the vector containing the basis functions and $\mathbf{B}_{W,k}^{(r)}$ is the current MCMC value of the column vector $\mathbf{B}_{W,k}$ in the MCMC (see \eqref{eq:mix_exp}).

\begin{figure}[!htbp]
    \centering
    \includegraphics[width = 0.66\textwidth]{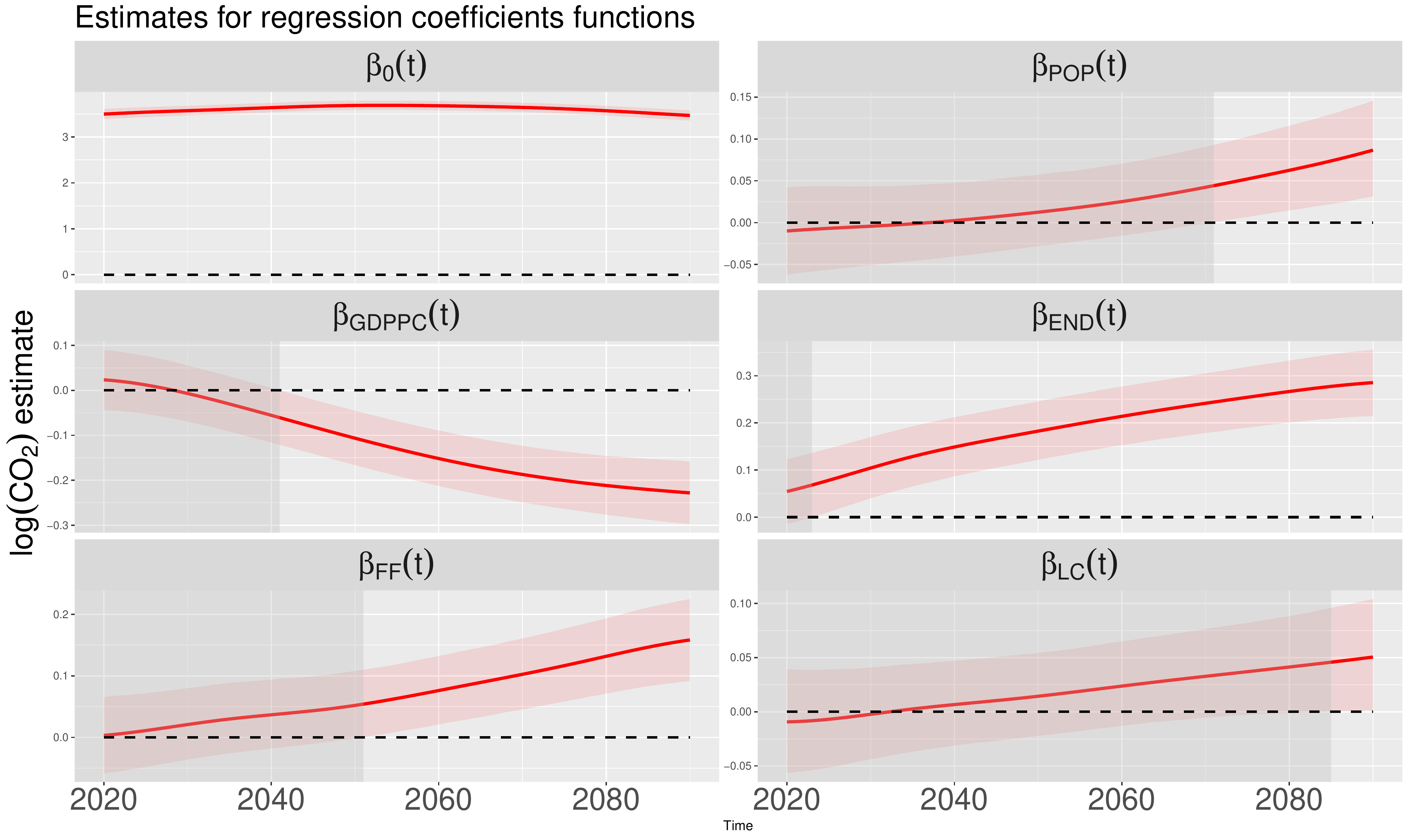}
    \caption{Estimates of the regression coefficient $\beta_{k}(t)$, obtained via B-spline basis expansion. The estimate is obtained as the posterior mean \eqref{eq:rand_est_coeffs} and denoted as solid red line.}
    \label{fig:fixrand_beta_est_plot_2}
\end{figure}

Figure~\ref{fig:fixrand_beta_est_plot_2} shows the estimated regression coefficient functions computed, as in \eqref{eq:rand_est_coeffs}, over a fine grid of values $t$ in the continuous interval $[2020,2090]$, along with their $95\%$ posterior credibile bands. The plots also show, in grey shading, the time points in which each $95\%$ credibility interval does include zero. This is a useful tool to interpret and assess the \lq\lq significance\rq\rq of the regression coefficients depending on time. More precisely, we consider the time interval in which zero is not included in the credibility band as the time interval in which the variable is significantly different from zero.

The continuous framework of model \eqref{eq:time_model}, through a functional data approach makes possible estimation, visualization and assessment of the regression coefficient functions. This is important since it allows to assess the time intervals in which the effect are strong and those in which such effect is null or negligible. Furthermore, the variation through time of each $\beta_{k}(t)$ is another important we are going to discuss here below.

The posterior credibile bands of coefficients corresponding to \textbf{GDPPC} and \textbf{END} exclude $0$ from about $2040$ and $2020$, respectively, and this gives evidence to consider them as the principal driving fixed effects in the IAMs simulations. In addition, they are also the coefficients more deviating from zero, indicating that their effect magnitude is quite large. Concerning the sign of the regression coefficient functions, \textbf{GDPPC} has a negative effect, while all the others have a positive effect. This negative effect can be explained since scenario SSP1 assumes that the world is richer than in SSP3. Typically a richer population will consume and pollute more than a poorer one. Covariates \textbf{FF} and \textbf{POP} become significant after $2050$ and $2070$ respectively, and, although the magnitude of the associated coefficients is not as big as the one relative \textbf{GDPPC} and \textbf{END}, they can also be considered significant ad well. Variable \textbf{LC} does not seem to be significant. See Section \ref{sec:reg_coeffs_sign} of the Supplementary Materials for an alternative assessment of the significance of the coefficients.

\subsection{Emission curve estimation}\label{subsec:em_curve_est}

\begin{figure}[!htbp]
    \centering
    \includegraphics[width = 0.66\textwidth]{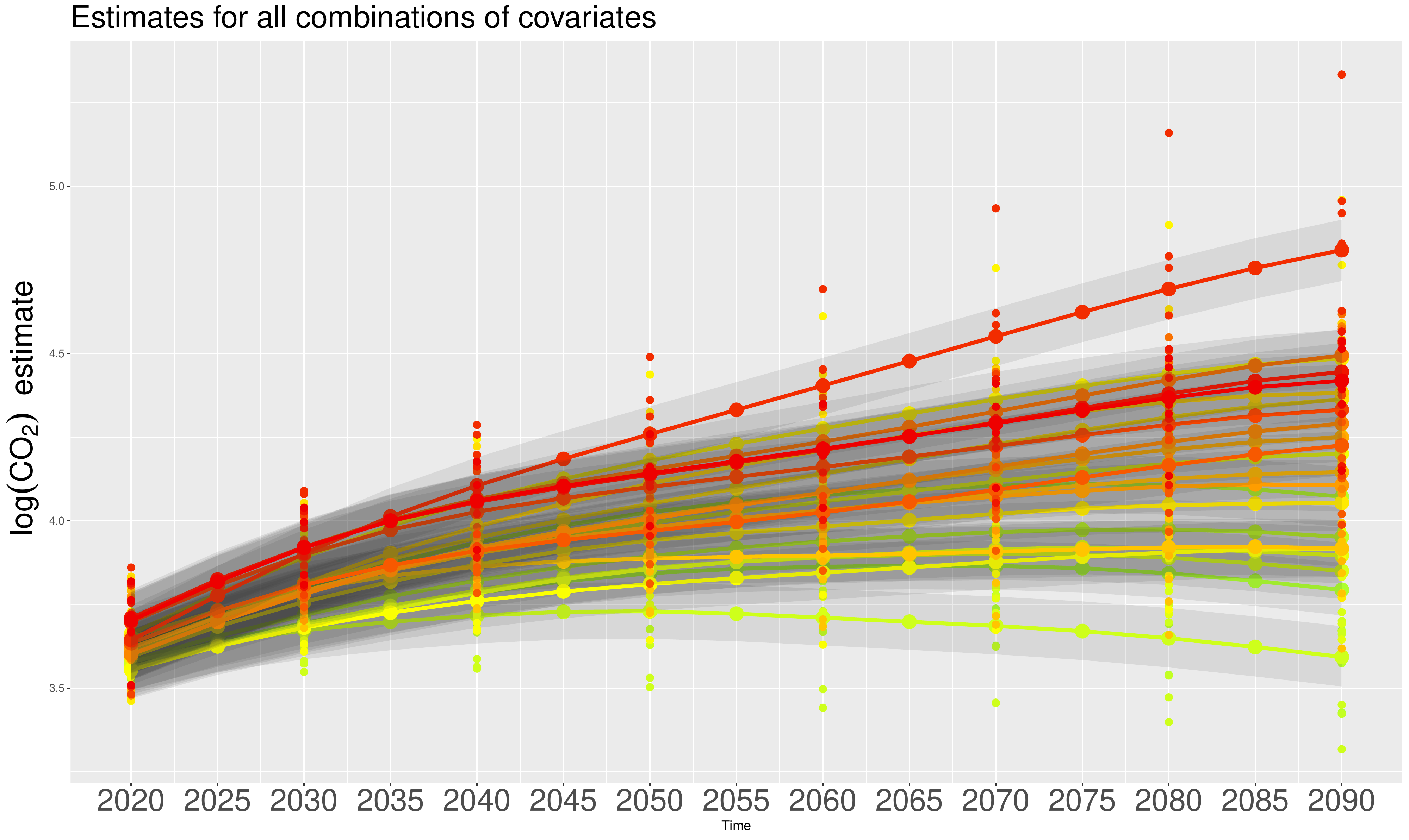}
    \caption{Posterior mean of $c_{i}(t)$ corresponding to the 23 combination of SSPs in the study with $95\%$ credibility bands associated; points are the observed outputs given by the IAMs.}
    \label{fig:fixrand_pred}
\end{figure}

As mentioned in the Introduction, one of the big advantages of having estimated, together with the other parameters, the expansion score matrix $\mathbf{B}_Z$, is that we now can estimate the parameters which represent the conditional expectation of $y_{ij}(t)$ in \eqref{eq:rand_eq}. We compute the posterior mean of the functional random effects parameters $c_{i}(t)$ in \eqref{eq:time_model} through representation \eqref{eq:rand_eq}. Specifically we compute $\mathbb{E}\bigl[c_{i}(t)|data\bigr]$ for $i=1,...,I$ through the MCMC output, similarly as in \eqref{eq:rand_est_coeffs} starting from posterior draws of $\mathbf{B}_{Z}$. This posterior estimate represent the Bayesian estimates of the logarithm of the CO\textsubscript{2} emission at any time point $t$ in the interval $[2020,2090]$. Figure~\ref{fig:fixrand_pred} displays the $c_{i}(t)$ posterior means along with the $95\%$ credibility intervals.

\begin{figure}[!htbp]
    \centering
    \includegraphics[width = 0.66\textwidth]{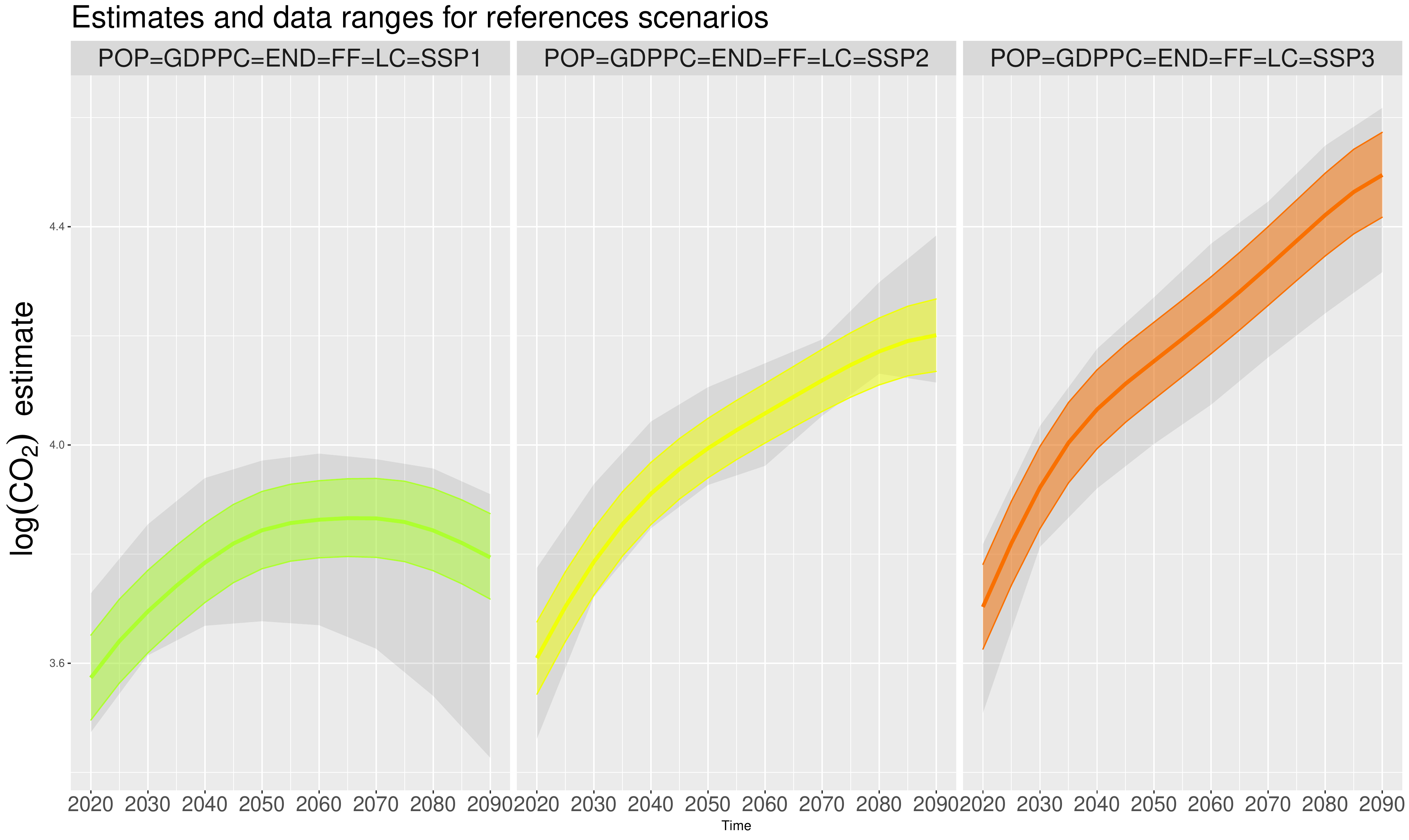}
    \caption{Mean estimation on the reference scenarios taking the same value in each SSP variable.}
    \label{fig:SSP_pred}
\end{figure}

To better understand the estimates in Figure~\ref{fig:fixrand_pred} we have plotted, in Figure~\ref{fig:SSP_pred}, for the SSPs combinations that take the same value in each SSP variable (i.e. $\mathbf{w}_{i_{1}} = (1,1,1,1,1,1)^T$, $\mathbf{w}_{i_{2}} = (1,2,2,2,2,2)^T$ and $\mathbf{w}_{i_{3}} = (1,3,3,3,3,3)^T$), the mean and the $95\%$ credible bands over time of the posterior distribution of $c_{i}(t)$ together with the min-max data range (i.e. the grey shading). Differently from the standard approach for uncertainty quantification of computer simulators, usually followed for the IAMs, here, in a probabilistic way, we account and give a measure for their uncertainty. In fact we have computed the full posterior distribution of the IAMs output mean given by SSPs specific input. This has enormous advantages in terms of interpretation of the estimates, since we can give true probabilistic interpretation of these estimates. Nowadays the uncertainty quantification of IAMs is carried out through an empirical evaluation of the output of different simulators.

From Figure \ref{fig:SSP_pred} it is clear that the popular empirical approach uses only deterministic ranges (the grey bands in Figure \ref{fig:SSP_pred}), which does not take into account any variability in the process. Our estimates, instead, use the proper posterior distribution on the emissions at unobserved data points, thanks to the functional framework. It is worth to specify that this is a novelty in the IAMs framework since, as we mentioned before, the uncertainty quantification of IAMs is principally carried out through the evaluation of the empirical ranges of different models. As a matter of fact, not only we are representing the uncertainty hidden into the multi-IAM framework, but, through the emulation of the IAM, we are also providing a probabilistic meaning to the parameters estimates, a novelty for this kind of simulations.

There was very little difference in terms of posterior predictive MSE under our model and the MSE derived through empirical estimates, computed following Chapter 13 of the book by \cite{ramsay2005functional}, since our model MSE is 0.016 and the frequentist one is 0.014. Although it is slightly better the frequentist one we are more satisfied with our model since it benefits from all of the Bayesian advantages that we have listed throughout the paper. In Supplementary Material, Section \ref{sec:images}, Figure \ref{fig:bayes_vs_freq} shows three estimates comparison between the two approaches.

\subsection{Temporal kriging}

\begin{figure}[!htbp]
    \centering
    \includegraphics[width = 0.66\textwidth]{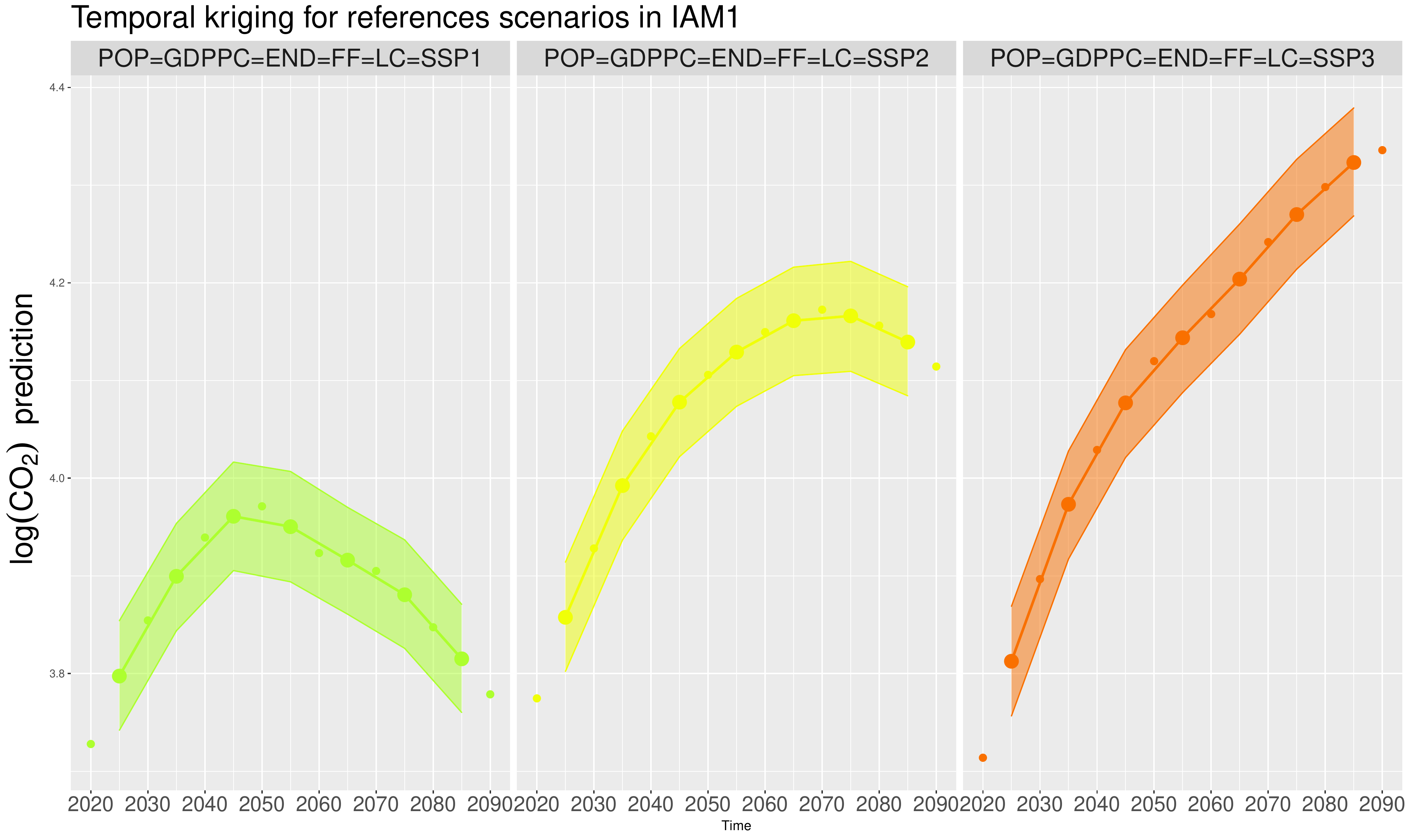}
    \caption{Temporal kriging for three reference scenarios of IAM 1}
    \label{fig:temporal_kriging}
\end{figure}

In order to properly perform predictive temporal downscaling, we adopt a kriging approach, tailored on our purposes. More precisely, here we want to predict the values of the emissions at any mid-decade (i.e, at years $2025$, $2035$, ...,$2085$) for each IAM and for each SSP combination. Then, we adopt an augmented data approach, so that we consider, as response variables:
\begin{equation*}
    \widetilde{\mathbf{y}}_{ij} = \bigl[y_{ij}(t_{1}),y_{ij}(t_{2}),...,y_{ij}(t_{\widetilde{D}})\bigr]
\end{equation*}
where, in this case, $t_{1} = 2020$, $t_{2} = 2025$, $t_{3} = 2030$, $t_{4} = 2035$, ..., $t_{\widetilde{D}} = 2090$ with $\widetilde{D} = 15$. The model for these augmented data is exactly the same as in \eqref{eq:rand_eq} where matrices $\widetilde{\bm{\Theta}}$ and $\widetilde{\bm{\Sigma}}$ are defined as $\bm{\Theta}$ and $\bm{\Sigma}$, respectively, with the proper changes of dimensions. Specifically, we assume:
\begin{equation*}
    \widetilde{\mathbf{y}}_{ij}| \mathbf{B}_{Z}, \widetilde{\bm{\Sigma}} \stackrel{ind}{\sim} \mathcal{N}_{\widetilde{D}}(\mathbf{z}^{T}_{i} \mathbf{B}^{T}_{Z} \widetilde{\bm{\Theta}}^{T},\widetilde{\bm{\Sigma}})
\end{equation*}
Here we distinguish between observed emissions at full decades, and non-observed data, i.e., the emissions at years $2025$, $2035$, ...,$2085$. In case of missing observations, the Bayesian approach allows them to be treated as (random) parameters. In practice, we include them in the state space of the MCMC and simulate from their posterior predictive distribution. Specifically, we draw MCMC samples from the following distribution, i.e.
\begin{equation}\label{eq:pred_dist}
    \mathcal{L}(\mathbf{y}^{\text{pred}}_{ij}|\mathbf{y}^{obs}) = \int_{\Phi} \mathcal{L}(\mathbf{y}^{\text{pred}}_{ij},d\phi|\mathbf{y}^{obs}) = \int_{\Phi} \mathcal{L}(\mathbf{y}^{\text{pred}}_{ij}|\phi,\mathbf{y}^{obs}) \pi(\phi|\mathbf{y}^{obs})d\phi
\end{equation}
where $\mathbf{y}^{\text{pred}}_{ij} = \bigl[ y_{ij}(t_{1}),y_{ij}(t_{3}),...,y_{ij}(t_{\widetilde{D}}) \bigr]$, $\mathbf{y}^{\text{obs}}$ contains all the observed data, $\pi(\phi|\mathbf{y}^{obs})$ is the posterior of the vector of all the parameters of the model and $\Phi$ is the space of all the model parameters. The distribution $\mathcal{L}(\mathbf{y}^{\text{pred}}_{ij}|\phi,\mathbf{y}^{obs})$ can be computed from the joint distribution of $\mathbf{y}^{\text{pred}}$ and $\mathbf{y}^{\text{obs}}$, which is a $\widetilde{D}$-dim Gaussian density. From straight-forward computations we derive that
\begin{equation*}
    \mathbf{y}^{pred}_{ij} | \mathbf{B}_{Z}, \widetilde{\bm{\Sigma}}, \mathbf{y}^{obs}_{ij} \sim \mathcal{N}_{D_{pred}}(\Bar{\mathbf{\mu}}_{ij}, \Bar{\bm{\Sigma}})
\end{equation*}
where
\begin{align*}
    \Bar{\mathbf{\mu}}_{ij} &= \bm{\Theta}_{pred} \mathbf{B}_{Z} \mathbf{z}_{i} + \bm{\Sigma}_{pred,obs} \bm{\Sigma}_{obs}(\mathbf{y}^{obs}_{ij}-\bm{\Theta}_{obs} \mathbf{B}_{Z} \mathbf{z}_{i}) \\
    \Bar{\bm{\Sigma}} &= \bm{\Sigma}_{pred,pred} - \bm{\Sigma}_{pred,obs} \bm{\Sigma}^{-1}_{obs,obs} \bm{\Sigma}_{obs,pred}
\end{align*}
with $\bm{\Sigma}_{pred,pred}$, $\bm{\Sigma}_{obs,obs}$, $\bm{\Sigma}_{pred,obs}$ and $\bm{\Sigma}_{obs,pred}$ are the matrices relative to the predicted points covariance, to the observed points covariance, and to the covariance between observed and predicted points. Moreover, $\bm{\Theta}_{pred}$ is the submatrix of $\widetilde{\bm{\Theta}}$ relative to the predicted points respectively. Figure~\ref{fig:temporal_kriging} displays, for three curves of IAM 1, the distribution in \eqref{eq:pred_dist}, i.e. the temporal kriging, represented by bigger dots and the shades (the $95\%$ credible band), and the associated observed data by smaller dots.

The plot shows that the IAM observed data lie inside the $95\%$ credibility intervals of the estimated mean curves. Recall that, in general, computer models, such as the IAMs, are very expensive to run, and hence, they are ran only with a small amount of input points. In this perspective, our procedure is an interesting tool that can help refining the temporal grid, in order to lighten up the computational burden of the simulators. Moreover, it also allow researchers to draw conclusions based on few run multiple IAMs giving as output projections with decade frequencies.

\section{Conclusions}\label{sec:conclusions}

Climate change is one of the most difficult challenges humanity has ever faced \citep{pachauri2014climate}. Decreasing greenhouse gases emissions, especially CO\textsubscript{2}, is the most impacting action that, as a global society, we can do. For this reason, several computational models have been proposed by the scientific community in recent years, in order to simulate CO\textsubscript{2} emission profiles over this century, or other climatic variables. These simulators are complex models, which, generally, are very expensive to run. This implies that only a small number of runs, each with few different input parameters, can be performed. In the last decades, statistical emulators, i.e., statistical models, have gained attention as they emulate the behaviour of a simulator, using a few numerical outputs as data points to build inference.

Our work propose a Bayesian hierarchical model for simulator outputs. The model allows great flexibility and randomness in the model parameters. The Bayesian approach intrinsically offer a tool for uncertainty quantification, that is very important in this context. In fact, by computing the joint posterior distribution of all parameters, we are able to compute also interval estimates based on a probabilistic model. In this way, we quantify the posterior probability that a parameter $\theta_{l}$ belongs to an interval $(a,b)$. The continuous representation of the IAMs output is obtained thanks to the functional framework. This modelling choice is is justified mainly because we assume that the IAM emissions are produced as "simulated" output of a smooth phenomenon in time and space and, hence, it is convenient to treat them as a continuous function of time. Finally, the hierarchical structure of the model allows for more flexibility by introducing new parameters other than the global ones. As a matter of fact, in order to deal with nested data we needed group-specific parameters and hierarchical priors to share information between different groups to help parameter estimation. This is the so called "borrowing of informations" of these kind of Bayesian models. 

We have assumed a $AR(1)$ covariance structure for the likelihood. This choice is particularly useful to decrease the computational burden and increase the ability of parameters interpretation. Other modeling choices could have been adopted, such as time series models or multivariate regression. However, $AR(1)$ covariance structure is particularly convenient when considering temporal downscaling. 

The posterior analysis has showed that the most important factors are the \textbf{GDPPC} and \textbf{END}. This result emphasizes the need of far-sighted and \lq\lq integrated\rq\rq policies that includes changes,with respect to what has been done until now, in crucial fields such as the economic and energetic policies. We have also found that the the availability of fossil fuel is, statistically significant to predict the CO\textsubscript{2} emissions, although less important than other factors. In addition, posterior evidence shows that the variable describing the development of low carbon technologies does not influence the CO\textsubscript{2} simulated emissions almost at all. This may indicate that the decrease of the consumption and any change in western lifestyle can impact than the development of low carbon technologies in order to mitigate climate change. Finally, this work underlines the need of wide multi-disciplinary policy strategies that include all of the variables that have been found to be statistically significant. It stands out that it is fundamental to concentrate the mitigation efforts on the key drivers of climate change, and that without a broad strategy, a great cut of the emission is much more difficult to achieve.

In conclusion, deterministic models, that produce forecasts coming from different research group, are important tools that could allow policy makers to better understand the processes they are called to make decisions on. As a matter of fact, none of the existing IAMs is a perfect model for future projections, but each of them capture different key features of the phenomenon. In this work we have modeled the intrinsic variability when considering several IAMs outputs with identical inputs. This variability represents the uncertainty that the different modelling assumptions and settings induce on the CO\textsubscript{2} emission processes.

In the end, we believe that we have provided a new insight into the Bayesian uncertainty quantification of computer simulated (i.e. via the use of IAMs) future projections of climate variables such as CO\textsubscript{2} emissions. This is crucial because, not only it allows to have uncertainty intervals with a probabilistic meaning, but also immediate interpretation of the results that are obtained. Our work built a model able to give a unified IAMs framework to the policy makers, giving the chance to climate scientist to emulate several computer models at once, with one statistical emulator. We did this by taking advantage of the progresses made in Bayesian inference for functional data, in completely unrelated applications compared to ours.

\bibliographystyle{rss}
\bibliography{biblio}

\clearpage

\pagenumbering{arabic}

\end{document}


\maketitle

\section{The Choice of Hyper-parameters and predictive comparison}\label{sec:hp_choice}

We use here the empirical Bayes approach to select weakly informative prior in \eqref{eq:rand_prior} in the manuscript. We follow the idea described in \cite{goldsmith2016assessing}. The values are chosen according to the following formulas:
\begin{align*}
    a_{Z} =& \frac{I \cdot K}{2} & b_{Z} &= \frac{1}{2} \text{tr}\biggl\{\bigl[(\mathbf{B}^{OLS}_{Z})^T - \mathbf{W}(\mathbf{B}^{WLS}_{W})^T\bigr]^T \mathbf{P}\bigl[(\mathbf{B}^{OLS}_{Z})^T - \mathbf{W}(\mathbf{B}^{WLS}_{W})^T\bigr] \biggr\} \\[1em]
    a_{W,k} =& \frac{K}{2} & b_{W,k} &= \frac{1}{2} (\mathbf{B}_{W,k}^{WLS})^T\mathbf{P}(\mathbf{B}_{W,k}^{WLS})
\end{align*}
where $\mathbf{B}^{OLS}_{Z}$ and $\mathbf{B}^{WLS}_{W}$ are respectively the frequentist \textit{Ordinary Least Squares} and \textit{Weighted Least Squares} estimates obtained by
\begin{align*}
    &\underset{\mathbf{B}_Z}{\text{min}} \text{\,\,\,} (\mathbf{Y} - \mathbf{Z} \mathbf{B}_Z^T \bm{\Theta}^T)^T(\mathbf{Y} - \mathbf{Z} \mathbf{B}_Z^T \bm{\Theta}^T) \\[1em]
    &\underset{\mathbf{B}_W}{\text{min}} \text{\,\,\,} (\mathbf{B}_Z - \mathbf{W} \mathbf{B}_W^T)^T \mathbf{P} (\mathbf{Y} - \mathbf{W} \mathbf{B}_W^T)
\end{align*}
Different choices of hyper-parameters will be compared via predictive goodness of fit indexes.

For model \eqref{eq:rand_eq}-\eqref{eq:arh1_prior} described in the manuscript we have compared three different choices of the hyper-parameters. More precisely, we compared the following configurations

\begin{center}
    \begin{tabular}{ c|c|c|c } 
        & \textbf{Reference} & \textbf{Alternative 1} & \textbf{Alternative 2} \\
        \hline
        $a_{W,k}$ & $4$ & $3$ & $5$ \\ 
        \hline
        $\mathbf{b}_{W}$ & $[0.51,0.0002,0.002,0.001,0.0005,0.0001]^T$ & $[2,3,4,5,6,7]^T$ & $[2,4,6,8,10,12]^T$ \\
        \hline
        $a_{Z}$ & $92$ & $2$ & $4$ \\ 
        \hline
        $b_{Z}$ & $0.0038$ & $6$ & $7$ \\
        \hline
        $\nu$ & $7$ & $11$ & $20$ \\
    \end{tabular}
\end{center}    
using two well known predictive goodness of fit indexes, the widely applicable information criterion \citep[WAIC][]{watanabe2013waic} and the log pseudo marginal likelihood \citep[LPML][]{geisser1979predictive}.

\begin{center}
    \begin{tabular}{c|c|c|c}
         & \textbf{Reference} & \textbf{Alternative 1} & \textbf{Alternative 2} \\
        \hline
        LPML & 1362.296 & 1257.146 & 1255.931 \\
        \hline
        WAIC & -2534.813 & -2642.834 & -2642.104
    \end{tabular}
\end{center}

Recall that the best model is the one with greatest LPML and with the smallest WAIC. From the LPML it seems that the reference set of the hyper-parameters is the best choice while WAIC suggests that one of the alternative choices is better. Interpreting predictive performance indexes is not an easy task but in this situation we can assume that our model is stable with respect to the choice of the hyper-parameters.

\newpage

\section{Images}\label{sec:images}

\begin{figure}[!htbp]
    \centering
    \includegraphics[width=0.75\textwidth]{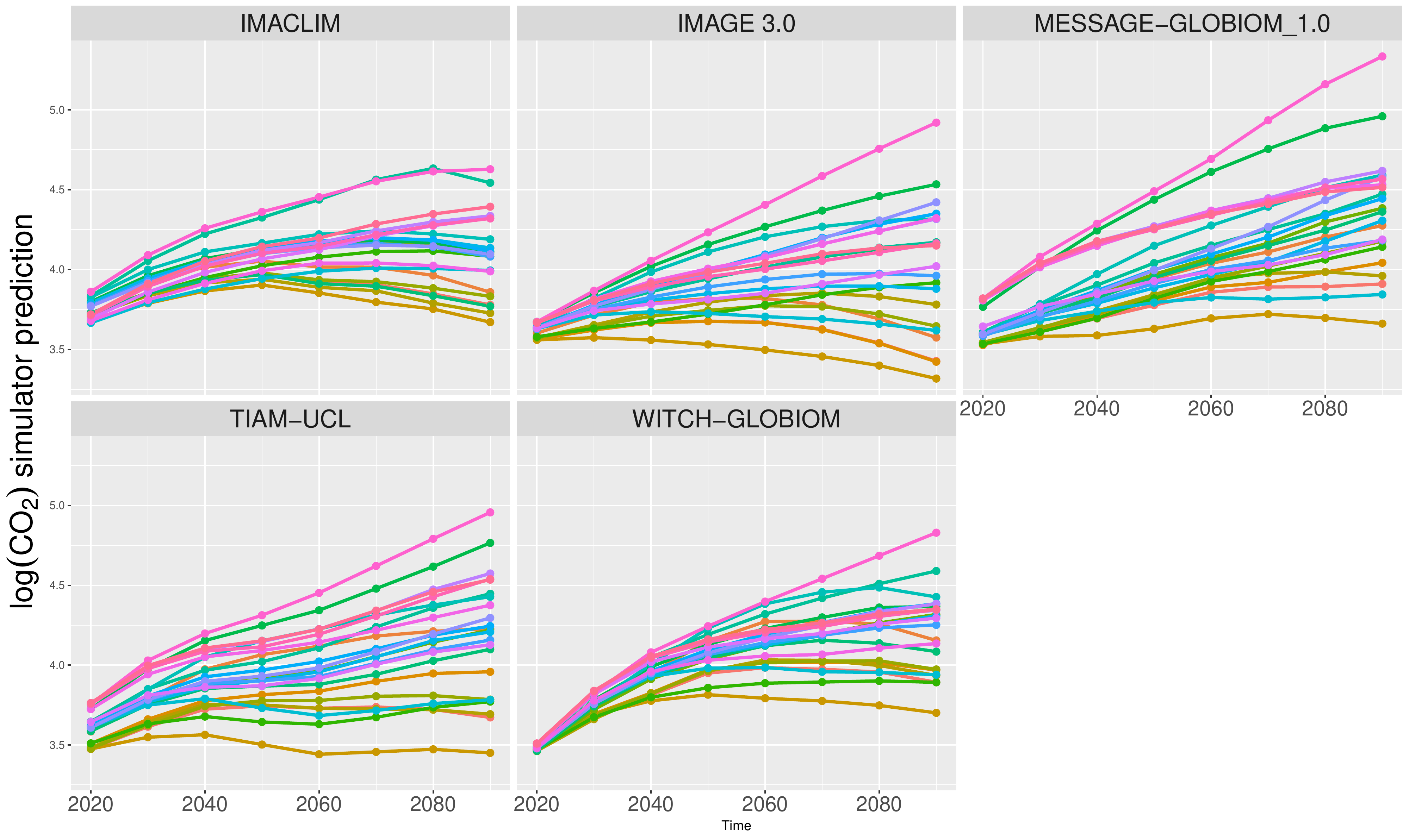}
    \caption{CO\textsubscript{2} emissions profiles in logarithmic scale grouped by IAM; each panel displays the 23 output from the same IAM.}
    \label{fig:models_t_ser}
\end{figure}

\begin{figure}[!htbp]
    \centering
    \includegraphics[width = 0.85\textwidth]{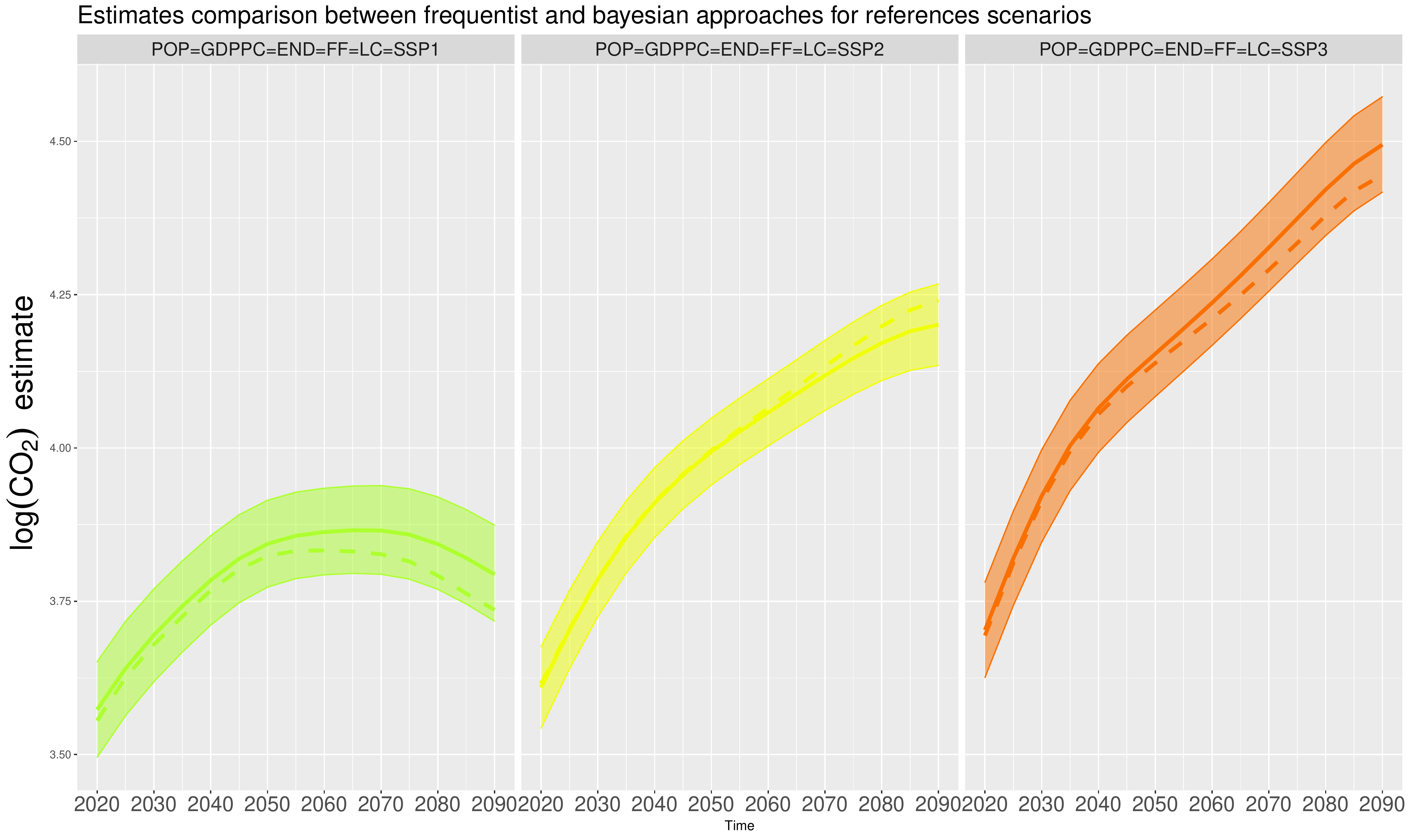}
    \caption{Comparison between Bayesian estimate (solid line + CI band) and the frequentist estimate (dashed line)}
    \label{fig:bayes_vs_freq}
\end{figure}

\begin{figure}[!htbp]
    \centering
    \includegraphics[width = 0.9\textwidth]{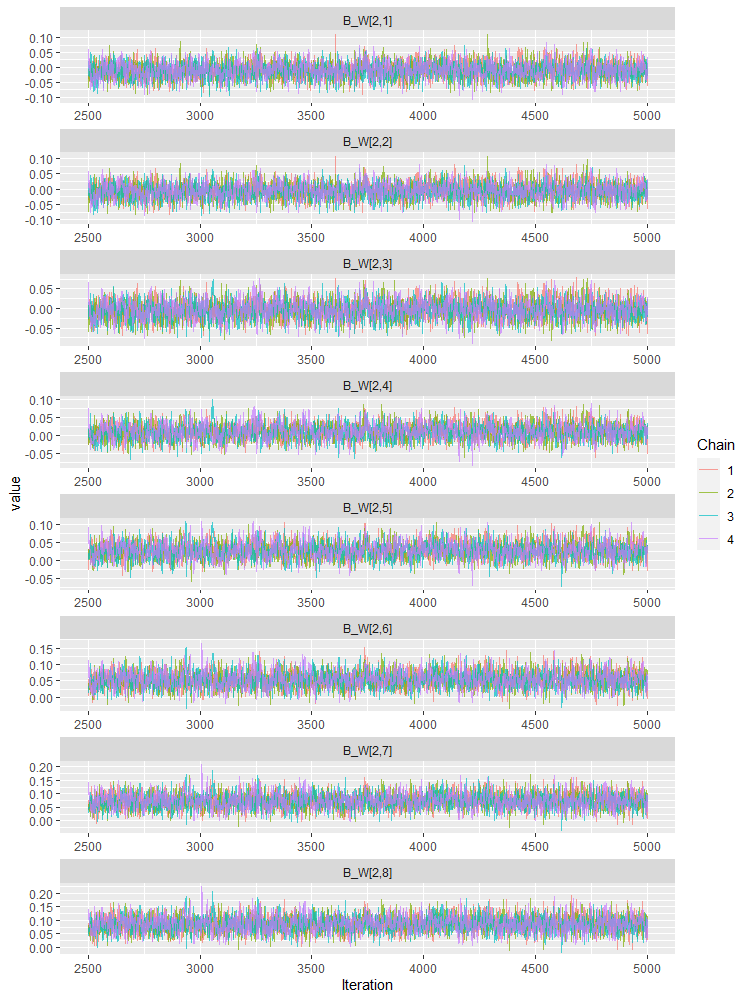}
    \caption{Traceplots of the regression coefficient expansion scores regarding POP variable}
    \label{fig:traceplot_POP}
\end{figure}

\begin{figure}[!htbp]
    \centering
    \includegraphics[width = 0.9\textwidth]{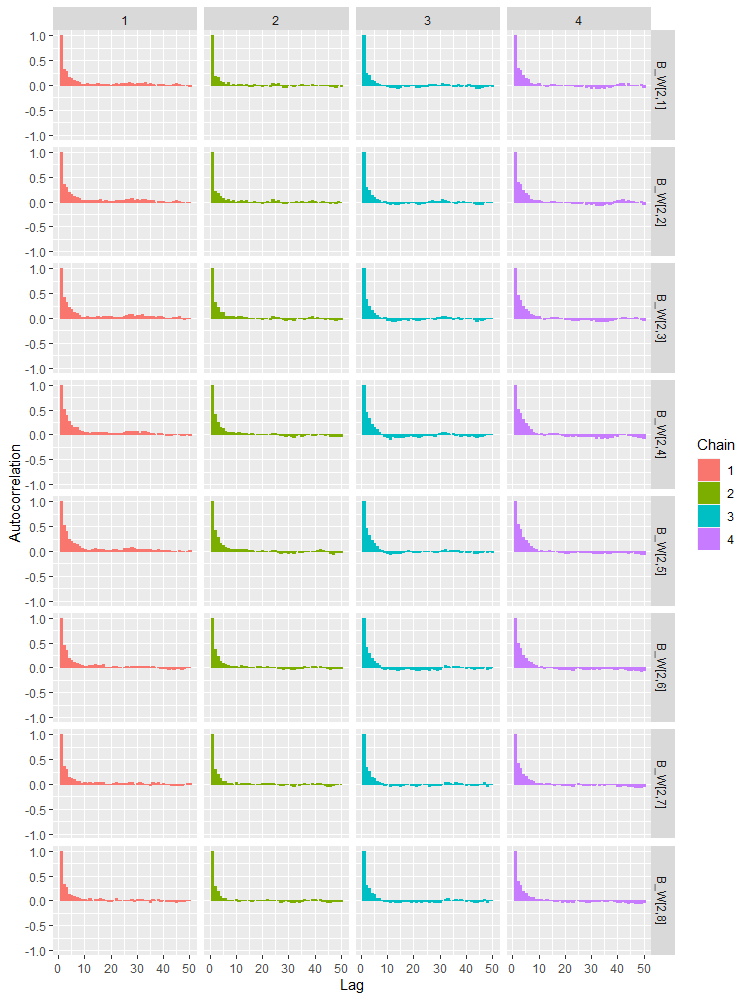}
    \caption{Autocorrelation of the regression coefficient expansion scores regarding POP variable}
    \label{fig:autocorrelation_POP}
\end{figure}

\newpage

\section{Regression coefficient function significance}\label{sec:reg_coeffs_sign}

In order to give a statistical evaluation of the significance of the regression coefficient functions, other than the evaluation of the credible bands over time, we compute the posterior probability over time of $\beta_{k}(t)$ being \lq\lq significantly\rq\rq different from $0$. This is another way to assess the estimates shown in Figure \ref{fig:fixrand_beta_est_plot_2} of the manuscript. 

Of course computing the posterior probability that $\beta_{k}(t) \neq 0$ is pointless because $\beta_{k}(t)$ has a continuous posterior distribution. \cite{kruschke2018bayesian} proposed the use of a \textit{region of practical equivalence} (ROPE) around the interested value (in this case $0$). In this way every time a value falls inside such interval, it can be considered equal to the value on which the interval is centered. For the purposes of this work, we consider, for each $t$ and $k=0,1,...,p$, an arbitrarily small interval centered around $0$ such as $(0 - \delta_k(t), 0 + \delta_k(t))$, with $\delta_k(t)$ equal to the posterior variance of $\beta_{k}(t)$. Because of \eqref{eq:mix_exp} from the main manuscript we have:
\begin{equation*}\label{eq:beta_vars}
   \delta_k(t) = \Var\bigl[\beta_{k}(t)|rest\bigr] = \Var\bigl[\bm{\Theta}(t)\mathbf{B}_{W,k}|rest\bigr] = \bm{\Theta}(t) \Var\bigl[\mathbf{B}_{W,k}|rest\bigr]\bm{\Theta}(t)^T
\end{equation*}
where $ \Var\bigl[\mathbf{B}_{W,k}|data\bigr]$ can be estimated from the posterior MCMC. 

\begin{figure}[!htbp]
    \centering
    \includegraphics[width = \textwidth]{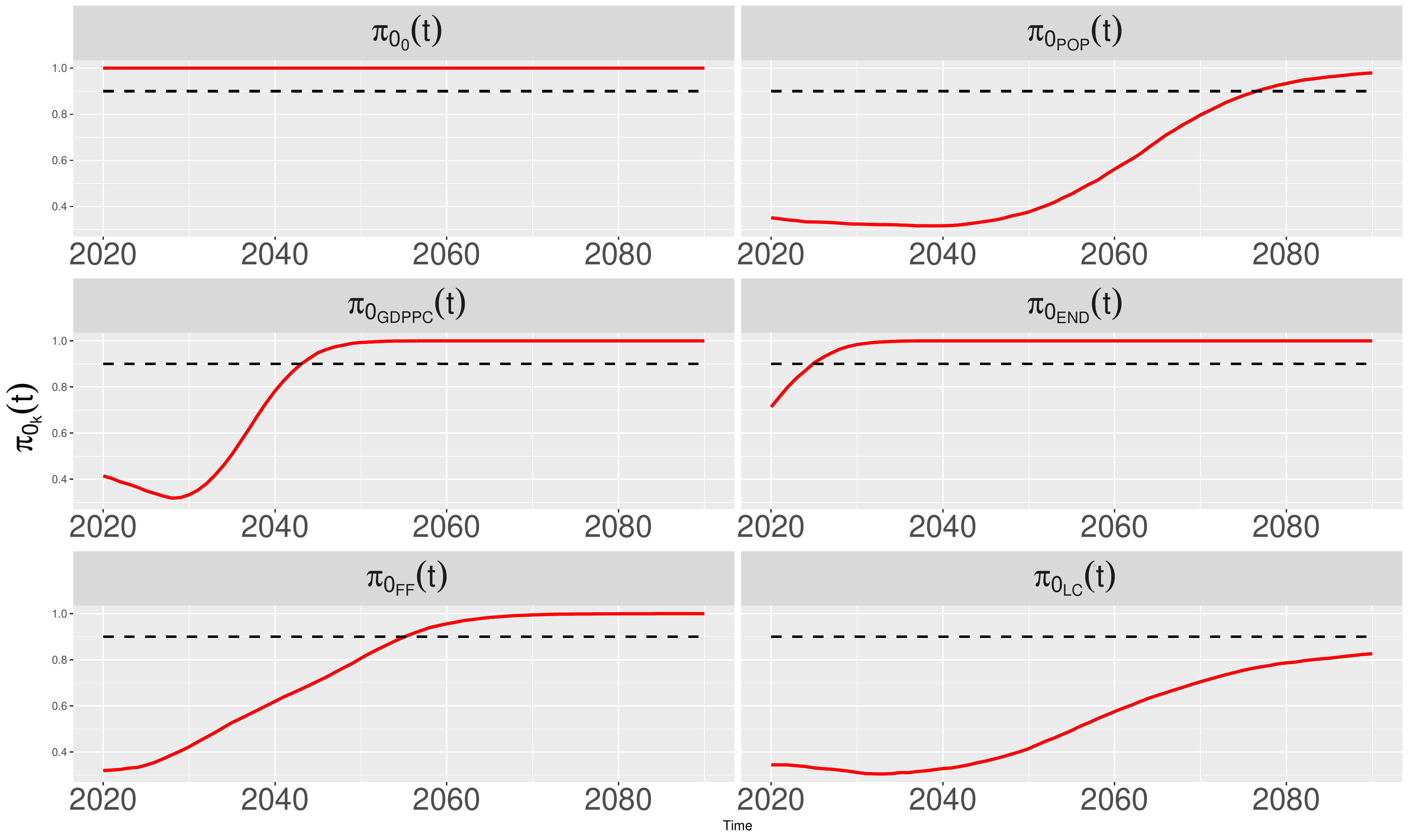}
    \caption{Plot of the $\pi_{0,k}(t) = \mathbb{P}(|\beta_k(t)|>\delta_{k}(t))$ (in solid red) referring to $\beta_{k}(t)$ for each $k=0,1,...,5$.}
    \label{fig:fixrand_coeff_imp}
\end{figure}

Therefore, for $t \in \{2020,2021,...,2090\}$ and $k = 0,1,...,p$, Figure \ref{fig:fixrand_coeff_imp} displays the following probability:
\begin{equation}\label{eq:imp}
    \pi_{0,k}(t) = \mathbb{P}\bigl(|\beta_{k}(t)| > \delta_k(t)|data\bigr) 
\end{equation}
for $k=0,1,...,p$. Following the ROPE principle, the computation of this probability is \textit{practically equivalent} to the probability of the coefficients of being different from zero.

The black dotted line at level $0.9$ denotes the threshold set to assess the significance of the variables. Coherently with the conclusion drawn from Figure \ref{fig:fixrand_beta_est_plot_2} of the manuscript we see that \textbf{GDPPC} and \textbf{END} are the variables that have the longest influence on the output of the model. They have similar behaviour, confirmed by the corresponding $\pi_{0,k}(t)$, with opposite effects as we have shown (respectively negative and positive). The first one surpass $0.9$ about in $2040$ while the second does it almost immediately after $2020$. Concerning \textbf{FF}, it is not as fast as the first two in exceeding $0.5$ and it surpass $0.9$ around $2050$. Although $\beta_{FF}(t)$ does not deviate from zero as much as \textbf{GDPPC} and \textbf{END}, it is clearly different from zero after 2050. These considerations on SSP variables are, somehow, analogous to those obtained in a frequentist framework by \cite{fontana2019functional}.

\section{Stan Code}\label{sec:codes}

\begin{lstlisting}[language = Stan]
functions { 
  matrix cov_matrix_ar1_obs(real ar, real sigma, int nrows) { 
    matrix[nrows, nrows] mat; 
    vector[nrows - 1] gamma; 
    mat = diag_matrix(rep_vector(sigma, nrows)); 
    for (i in 2:nrows) { 
      gamma[i - 1] = pow(ar, 2*(i - 1)-1); 
      for (j in 1:(i - 1)) { 
        mat[i, j] = sigma*gamma[i - j]; 
        mat[j, i] = mat[i, j]; 
      } 
    } 
    return mat; 
  }
  matrix cov_matrix_ar1_full(real ar, real sigma, int nrows) { 
    matrix[nrows, nrows] mat; 
    vector[nrows - 1] gamma; 
    mat = diag_matrix(rep_vector(sigma, nrows)); 
    for (i in 2:nrows) { 
      gamma[i - 1] = pow(ar, (i - 1)); 
      for (j in 1:(i - 1)) { 
        mat[i, j] = sigma*gamma[i - j]; 
        mat[j, i] = mat[i, j]; 
      } 
    } 
    return mat; 
  }
} 

data {
  int<lower=0> I;       // number of SSPs
  int<lower=0> J;       // number of IAMs for each SSP
  int<lower=0> IJ;      // total number of observations
  int<lower=0> p;       // number of fixed effects
  int<lower=0> Kt;      // number of spline basis functions
  int<lower = 0> D_obs;
  int<lower = 0> D_pred;
  int<lower = 1, upper = D_obs + D_pred> d_obs[D_obs];
  int<lower = 1, upper = D_obs + D_pred> d_pred[D_pred];
  vector[p] W[IJ];      // fixed effect design matrix
  vector[I] Z[IJ];      // random effect design matrix
  vector[D_obs] Y_obs[IJ];  // outcome matrix
  matrix[D_obs + D_pred,Kt] THETA;  // B-spline evaluation matrix
  cov_matrix[Kt] PenMat;    // prior penalty matrix for spline effects
  real<lower=0> aw;         // first hyper-parameter for the prior on sigma_B_W
  vector[p] bw;             // second hyper-parameter for the prior on sigma_B_W 
  real<lower=0> az;         // first hyper-parameter for the prior on sigma_B_Z
  real<lower=0> bz;         // second hyper-parameter for the prior on sigma_B_Z
  real<lower=0> v;          // hyper-parameter for the diagonal of covariance
}

transformed data {
  int<lower = 0> D = D_obs + D_pred;
  vector[Kt] mu_B_W;        // prior mean for spline effects
  for (k in 1:Kt) {
    mu_B_W[k] = 0;
  }
}

parameters {
  matrix[p,Kt] B_W;         // matrix of fixed effect spline coefficients
  matrix[I,Kt] B_Z;         // matrix of random effect spline coefficients
  vector<lower=0>[p] sig2_B_W;  // tuning variance
  real<lower=0> sig2_B_Z;       // tuning variance for random effect
  real<lower=0> sigma;          // residual variances
  real<lower=0> psi;
  real<lower=-1,upper=1> phi;   // autoregressive effects
}

transformed parameters {
  vector<lower=0>[p] tau2_B_W;  // tuning parameter for the variances of B_W
  real<lower=0> tau2_B_Z;       // tuning parameter for the variances of B_Z
  cov_matrix[D_obs] SIG;
  for(k in 1:p) {
    tau2_B_W[k] = pow(sig2_B_W[k], -1);
  }
  tau2_B_Z = pow(sig2_B_Z,-1);
  SIG = cov_matrix_ar1_obs(phi,sigma,D_obs);
}

model {
  // Prior for variance components controlling smoothness in beta
  for (k in 1:p) {
    sig2_B_W[k] ~ inv_gamma(aw,bw[k]); 
  }
  // Prior for variance components of the random effects
  sig2_B_Z ~ inv_gamma(az,bz);
  // Prior for spline coefficients for beta
  for (k in 1:p) {
    (B_W[k])' ~ multi_normal_prec(mu_B_W, tau2_B_W[k] * PenMat);
	}
  // prior for the spline coefficients for c 
	for (i in 1:I) {
    (B_Z[i])' ~ multi_normal_prec((B_W)'* W[i], tau2_B_Z * PenMat);
	}
  // Prior for covariance matrix of the residual
  phi ~ normal(0,1);
  sigma ~ inv_gamma(v/2,v*psi/2);
  psi ~ gamma(1,0.75);
	// Outcome likelihood
	for (n in 1:IJ) {
	  Y_obs[n, ] ~ multi_normal((THETA[d_obs,] * B_Z')*Z[n,], SIG);
  }
}

generated quantities {
  matrix[D_pred,IJ] Y_pred;   // augmented data vector including missing ones
  cov_matrix[D] SIG_full;
  SIG_full = cov_matrix_ar1_full(phi,sigma,D);
  for (n in 1:IJ)
    Y_pred[,n] = multi_normal_rng((THETA[d_pred,] * B_Z') * Z[n,] + SIG_full[d_pred,d_obs] * inverse_spd(SIG_full[d_obs,d_obs]) * (Y_obs[n,] - (THETA[d_obs,] * B_Z') * Z[n,]),SIG_full[d_pred,d_pred] - SIG_full[d_pred,d_obs] * inverse_spd(SIG_full[d_obs,d_obs]) * SIG_full[d_obs,d_pred]);
}

\end{lstlisting}

\bibliographystyle{apalike}
\bibliography{biblio}